# Percolation threshold and critical point of a nuclear reactor

V. V. Ryazanov

Institute for Nuclear Research, pr. Nauki, 47 Kiev, Ukraine, e-mail: vryazan19@gmail.com

Neutrons in a nuclear reactor move along trajectories corresponding to Cayley trees associated with branching random processes. The probability of percolation, the appearance of such a state of the Bethe lattice in which there is at least one continuous path through neighboring conducting nodes through the entire lattice, corresponds to the probability of the occurrence of a self-sustaining fission chain reaction. At a critical value of the last probability, a (conditionally) infinite cluster of neutrons appears. The probability of percolation, depending on the operating time of the reactor and its size, is associated with the criticality of the reactor. The temporal behavior of the neutron multiplication factor is estimated. Particular attention is paid to the early stages of the development of a self-sustaining chain reaction of nuclear fission. The possibilities of determining the boundaries of the critical region are indicated.

## 1. Introduction

The theory of percolation, which arose from problems of the flow of liquid or gas through a random labyrinth, has developed into a broad mathematical discipline with physical applications to magnetization, conductivity and other properties of various systems [1–6]. The theory of percolation describes the emergence of infinite connected structures (clusters) consisting of individual elements. Percolation is the moment when a lattice state appears in which there is at least one continuous path through adjacent conducting nodes through the entire lattice. The set of elements through which percolation occurs is called a percolation cluster. Percolation theory deals with the formation of bound objects in disordered media. From a mathematician's point of view, the theory of percolation should be classified as a theory of probability in graphs. From a physicist's point of view, percolation is a geometric phase transition.

Percolation phenomena are closely related to fractality, the phenomena of self-similarity and universality. Fractal models of various kinds of systems make it possible to discover new features of seemingly well-known phenomena. Many physical systems are fractal and multifractal. In [7], fractal is a structure consisting of parts that are in some sense similar to the whole. Fractal properties manifest themselves especially clearly at the very point of the phase transition, in the critical region. The steady-state operation of a nuclear reactor (NR) occurs precisely at the critical point, and the fractal description should be very important for characterizing the operation of the reactor.

In [2], the spread of rumors in the percolation model is compared with a chain reaction. The relations of the percolation theory [1, 4] are also valid in the general theory of phase transitions. Fractal concepts were used in the study of highly developed turbulence, inhomogeneous star clusters [8], diffusion-limited aggregation, processes of destruction of matter, the structure of blood, etc. The description of the physical properties of systems with a fractal structure led to the development of analytical methods in the fractal concept based on application of the mathematical apparatus of fractional order equations, since the dimension of space becomes fractional. The need to switch to neutron transfer equations in fractional derivatives may be of practical importance for reactor calculations [9], although there is often no sharp distinction between percolation processes and diffusion [4]. It was noted in [10] that transport processes in percolation clusters, fractal trees, and porous systems must be analyzed anew in order to obtain correct transport equations for such systems. In branching fractal structures, "super-slow" transfer processes can occur, when a physical quantity changes more slowly than the first derivative. The index of the fractional derivative with respect to time corresponds to the proportion of channels (branches) open to percolation. The dynamics of diffusion is determined by the random nature of particle motion: a diffusing particle can reach any point in the medium. Percolation is associated with a fractal medium: below the percolation threshold, the process of particle propagation is limited to a finite region of the medium. During diffusion from a source, a diffusion front appears that has a fractal structure. In [7], the term "shell" of a percolation cluster is introduced. Below, within the framework of the percolation theory, chain reaction processes in the reactor are considered.

The importance of the relations of the percolation theory for neutron processes in a reactor is clear from the fact that they allow one to immediately obtain the neutron multiplication equation and the equation of the critical size of the reactor, which interpret the general relations of the percolation theory, which indicates the effectiveness of this approach in the theory of neutron processes in a reactor. The relation for the speed of propagation of a disturbance at local supercriticality should prove useful. Many other expressions of percolation theory applied to reactors may also be of interest. This is apparently due to the fact that percolation is a critical process that presupposes the existence of a critical point, a certain threshold. At the threshold, flow occurs along a fractal set, the geometry of which is determined by criticality. The geometric characteristics of a fractal are independent of the microscopic properties of the medium. Below the critical point, kinetic processes are limited to a finite region of phase space, scattering, absorption and other neutron processes. At the critical point, the fractal set, which is formed when the free energy of the statistical ensemble decreases, becomes decisive. The behavior of the system under slow influences on it tends to self-organized criticality [11, 12]. Stationary nonequilibrium states on fractal structures are chaotic, turbulent in nature. In [13, 14], the Lorentz model is used to study them.

The kinetics and processes of transfer in fractal reactor structures require a separate detailed study [8, 9, 14, 15]. In the region of the critical point, long-range correlation effects appear, manifested in the non-Gaussian behavior of kinetic processes, determined by the topological invariants of self-similar fractal sets. Transfer processes at the percolation threshold are discussed in [8, 9]. Fractional derivative equations are used that take into account the effects of memory, nonlocality and intermittency.

There are few exact self-similarities in nature; in other words, fractals with constant dimensions are rare, but fractals with variable dimensions—multifractals—are common [4, 16]. Neutron processes in reactors are no exception.

The article shows that the criticality of a reactor depends not only on the probability of nuclear fission by a neutron and the number of neutrons produced after nuclear fission, but also on the number of generations of neutrons, depending on the time and size of the system. For breeding systems of small sizes (critical assemblies, small reactors), as well as for the reactor start-up stage, when the number of neutrons is small, these circumstances must be taken into account.

## 2. Relationship between the theory of percolation on Bethe lattices and the neutron multiplication factor. Results.

Division chains in NR have the geometric form of Cayley trees [1–4]. Analytical results can be obtained for Cayley trees. The percolation threshold on the Cayley tree is less than one. Therefore, regimes both below and above the percolation threshold can be considered.

A Cayley tree, also called a Bethe lattice, is a loopless structure that is constructed starting from a central node from which z branches of unit length emanate. They form the first shell of the Cayley tree. The end of each branch is also a node. z-1 new branches emanate from each node, forming $z(z-1)$ nodes of the second shell. The process continues ad infinitum. This produces an infinite Cayley tree with z branches emanating from each node. Any two nodes are connected by only one path. In this case, the random nature of branching should be taken into account. You can also apply random graph theory. Knowledge of the properties of clusters allows one to study their dynamic properties. There is a close connection between fractal phenomena and statistical distributions. The processes depicted by trees are associated with branching random processes [17], which describe neutron processes in a reactor [18]. We consider an aspect of the problem determined by the size and nature of the behavior of clusters - nodes connected to each other. By a node we mean a fissile nucleus (or a neutron introduced into the system— the root of a tree [19]); by a bond, we mean the trajectories of neutrons. Neutron absorption points form so-called hanging ends [19] (vertices of degree 1) or free ends.

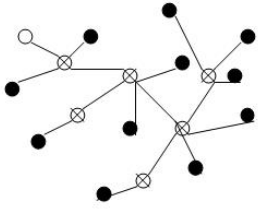

Fig. 1. Trajectories of neutrons and their descendants in the breeding medium: ○ - the point at which the initial neutron begins to move; ⊗ —points of fission of nuclei by neutrons; ● - neutron absorption points.

If it were possible to trace the trajectories of neutron movements in a nuclear reactor, then the observer would pay attention to the characteristic branching structure of the process for the total number of neutrons. In Fig. Figure 1 shows examples of the trajectories of one neutron introduced into a breeding medium, taking into account those evolutionary events (nuclear fission and neutron absorption) that lead to a change in the size of the neutron population.

At high subcriticality and large negative reactivity values $\rho=(k_{эф}–1)/k_{эф}$ (effective multiplication factor $k_{эф}\ll1$), the system contains small clusters with a predominant number of hanging ends. If the intensity of neutron death (absorption by the environment or leaving the system) during the time $\Delta t\to 0$ is denoted by $\lambda_c\Delta t+0(\Delta t)$, and the intensity of nuclear fission by a neutron $\lambda_f\Delta t+0(\Delta t)$ ($\lambda_f=v\Sigma_f$, where v is the neutron speed; $\Sigma_f$ is macroscopic fission cross section), then the probability of nuclear fission by a neutron is equal to

$$c=p=\lambda_f/(\lambda_f+\lambda_c). \quad (1)$$

Effective neutron multiplication factor $k_{ef}=p\bar{v}$, where $\bar{v}$ is the mathematical expectation of the number of secondary neutrons in one fission event. As $p$ increases, the cluster sizes increase. At $p=1$, all fuel nuclei in the nuclear reactor are separated, and $k_{ef\ max}=\bar{v}$ (under such conditions an explosion occurs). For $1–p\ll1$, there is an infinite cluster in the system. There must be a critical value $p_c$ at which a transition from one regime to another occurs—an infinite cluster appears for the first time. This corresponds to the case $k_{ef}=1$, $c_c=p_c=1/\bar{v}$. This result in the percolation model was obtained strictly mathematically [1–4, 7]. The formation of an infinite cluster represents a phase transition - the beginning of a self-sustaining chain reaction, the critical point of the system (in terms of reactor theory). An important role in the theory of phase transitions is played by the concept of an order parameter, that physical quantity that occupies a key place in the processes leading to the transformation. In the theory of percolation clusters, the order parameter is the power of an infinite cluster $P_\infty$ - the probability of a lattice site belonging to an infinite cluster. The critical behavior of this quantity at $p\to p_c$, $p>p_c$ is determined by the dependence

$$P_\infty=(p-p_c)^\beta, \quad (2)$$

where $\beta$ is one of the so-called critical indicators (scaling indices - in terms of percolation theory) [1, 4]. The value $\beta$ determines the critical behavior of the power of an infinite cluster $P_\infty$. In percolation theory, probability (2) is also called the percolation probability. It serves as the main characteristic of the percolation system. The percolation probability can be used to express properties of physical systems that depend on the topology of large clusters, such as spontaneous magnetization or conductivity. Such quantities as the average number of nodes of the final cluster, the correlation length $\xi$, the characteristic spatial scale of the cluster at $p<p_c$, and at $p>p_c$ - the characteristic size of the voids in it are also determined.

The formulas of the theory of percolation for the number of nodes and the correlation length in the theory of nuclear reactors (although there they are obtained in a different way) correspond to the equation for neutron multiplication $N=(1–k_{ef})^{-1}$, and the equation for the critical size $R_{эф}=\pi M(k_{ef}–1)^{-1/2}$, where $R_{ef}$ is the effective size, geometric parameter; M is the neutron migration length. In this case, the critical indicator is $\nu=1/2$. The use of percolation theory and constructions of fractal theory allows us to write down a number of other relations and consider, for example, dynamic critical indices, the dimension of the cluster skeleton, the spectral (fracton) dimension, etc. - many other properties strictly obtained for Cayley trees in [1], [3].

To study the multifractal properties of neutron processes in reactors, it is necessary to take into account the features of the fission process. If we construct the dependence of the multifractal spectrum $f(\alpha)$ [16], using the approach of works [4, 12], with a measure of the multiplicative population, we obtain the dependence presented in Fig. 2, a. The function for $f(\alpha)$ of the inhomogeneous Sierpinski triangle has a similar form [17]. The spectrum of generalized dimensions is also determined (Fig. 2, b) [16].

An important value of the percolation threshold is associated with probability $c=p$ (1). The set of elements through which flow occurs is called a percolation cluster. Being a connected random graph by nature, it can take different forms depending on the specific implementation. Therefore, it is customary to characterize its overall size. The percolation threshold is the number of elements of the percolation cluster divided by the total number of elements of the medium under consideration. Due to the random nature of switching states of the elements of the environment, in the finite system there is no clearly defined threshold (the size of the critical cluster), but there is a so-called critical range of values, into which the values of the percolation threshold obtained as a result of various random implementations fall with some probability. As the size of the system increases, the region narrows to a point

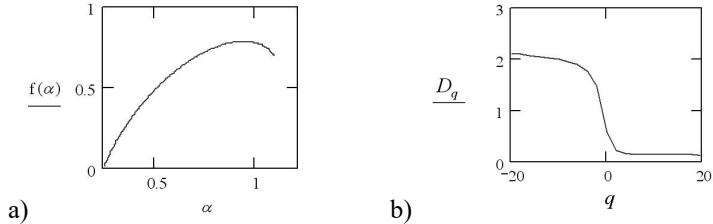

a) b)

Fig. 2. Function of the multifractal spectrum (a) and spectrum of generalized dimensions (b) for fission chains in a reactor taking into account delayed neutrons.

Processes on Bethe lattices are considered, as a rule, for the case of an infinite lattice. In this work, we consider the case of a finite lattice, which corresponds to a finite number of neutrons in the reactor. Taking into account the finite number of neutrons is necessary, for example, during reactor startup, in critical assemblies, in small reactors, etc. The results obtained can be useful for fast neutron reactors and for transient processes.

In addition to the probability of percolation and the percolation threshold, there are a large number of other characteristics of the percolation process [1].

For neutron processes in a nuclear reactor, the most important characteristics are the percolation probability, which is interpreted as the probability of a self-sustaining chain reaction, and the percolation threshold value, which is proportional to the neutron multiplication factor. In [20], a recurrence relation was obtained for the probability of percolation from the root vertex, the probability that a connected component of the configuration containing the root vertex (some starting point of the appearance of the first neutron in the system, which generated a chain reaction), reaches the opposite edges of the system. Conventionally, mathematically, the size of the system and the connected component tends to infinity, although real systems are finite. In [20], the value $P(n,c)$ denotes the probability of percolation from the root vertex to a distance n. The value n in our problem is interpreted as the number of generations of neutrons in a chain reaction. The number $c_{c\infty}=\inf\{c: P(c)>0\}$ is called the percolation threshold in [20]. In [4], this value is called the critical probability at which a cluster first appears, extending over the entire lattice. Here $P(c)=\lim_{n\to\infty}P(n,c)$, as in (2). In Fig., taken from [1], shows the behavior of the function $P(c=p)$. We will assume that for finite values of n there is a percolation threshold

$$c_{cn} = \inf \{c_n : P(n,c) > 0\}. \qquad (3)$$

In the general case, the probability of percolation has the form shown in Fig. 4 [1]. The recurrence relation obtained in [20] for the percolation probability has the form

$$P(n+1,c)=c[1-(1-P(n,c))^s]; \quad P(0,c)=c, \qquad (4)$$

where $s=\bar{\nu}$. In the continuous approximation there is a derivative of $P(n,c)$ with respect to $c$. To study the behavior of $P(n,c)$, for example, series expansion or determination of the inflection point, one must also know the second derivative. The expressions for derivatives obtained from (4) have the form

$$f(n,c)=dP(n,c)/dc \; ; \quad f(n+1,c)=1-(1-P(n,c))^{s-1}[1-P(n,c)-csf(n,c)]; \quad f(0,c)=1, \qquad (5)$$

$$r(n+1,c)=s(1-P(n,c))^{s-2}[2(1-P(n,c))(dP(n,c)/dc)-c(s-1)(dP(n,c)/dc)^2+c(1-P(n,c))r(n,c)];$$
$$r(n,c)=d^2P(n,c)/dc^2; \quad r(0,c)=0. \qquad (6)$$

From (4) we obtain a picture of the behavior of the percolation probability for the Bethe lattice (Fig. 4), which differs from that shown in Fig. 3, obtained for simple lattices. Probability $P(n,c)$ in Fig. 4 calculated at $n=750$. The vertical line shows the value $c_c=\bar{\nu}^{-1}$ at $n\to\infty$.

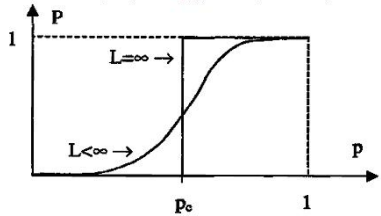

Fig. 3. Probability of percolation occurrence $P$ depending on the fraction of filled nodes $p=c$ (smooth curve corresponds to a lattice of finite size, stepped curve to an infinitely large lattice).

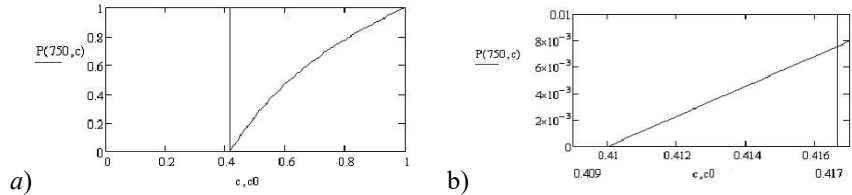

Fig. 4. Probability of percolation for the Bethe lattice, $n=750$ (ranges of variation with - 0…1 (*a*) and 0.409…0.417 (*b*)).

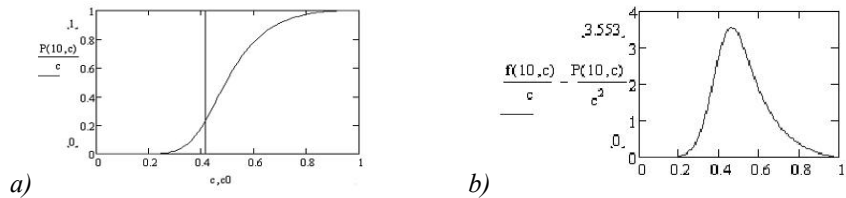

Fig. 5. Behavior of the conditional probability $P(10,c)/c$ (a) and the derivative with respect to $c$ of the function $P(10,c)/c$ (b).

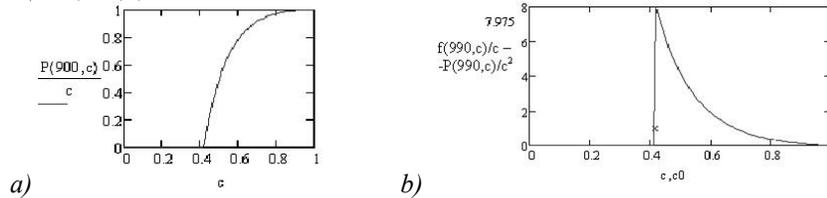

Fig. 6. Behavior of the function $P(900,c)/c$ (a) and the derivative with respect to $c$ of $P(990,c)/c$ (b).

Since the system is finite, the number of generations is $n=750$, the critical probability is not equal to $c_c=\bar{\nu}^{-1}$. This can be seen from Fig. 4, b, which shows a different range of changes in the value of $c$ - from 0.409 to 0.417, and not from 0 to 1, as in Fig. 4, a. From Fig. 4, b it is clear that the critical probability value for $n=750$ is less than for an infinite lattice, when $c_c=\bar{\nu}^{-1}$, $c_{c750}\approx 0{,}41<\bar{\nu}^{-1}$. In the interval from $c=0.41$ to $c_{c\infty}$ there is a non-zero probability of about $10^{-3}\ldots 8\cdot 10^{-3}$ of percolation or, for reactors, the probability of a self-sustaining chain reaction of fission of uranium nuclei. Thus, for finite systems the percolation threshold and the multiplication factor are less than unity.

Fig. 3 differs from Fig. 4. In Fig. 3 (a similar figure, which also shows the derivative $dP(n,c)/dc$, is in [3]) shows the conditional probability $P(n,c)/c$, corresponding in our interpretation to the fact that the initial neutron enters the fissile nucleus . In Fig. 5 shows the behavior of the function $P(10,c)/c$ and the derivative with respect to *c* of this function, similar to Fig. 3 and the figure from [3]. But this kind of behavior is typical for small values of *n*. As *n* increases (Fig. 6), the picture is more reminiscent of Fig. 4 and Fig. 7 for the derivative with respect to *c* of *P(n,c)*.

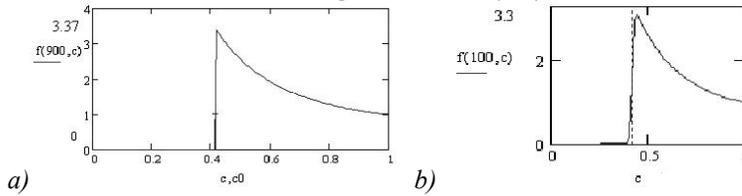

*a)* *b)*
Fig. 7. Dependence of the derivative with respect to *c* on *P(n,c)*, *n*=900, *a*) and *n*=100, b).

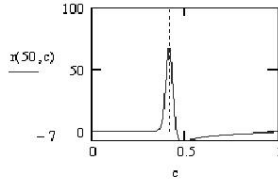

Fig. 8. Behavior of the second derivative with respect to *c* of *P(n,c)*, *n*=50.

The behavior of functions (5) and (6) has the form shown in Fig. 7, 8. It can be seen that the form of these functions is also related to the position of the percolation threshold, shown by the vertical dashed line.

In Fig. 9 shows the dependence of $P(n,c_0)$ on *n* for a fixed value $c=c_0=\overline{V}^{-1}=c_{c\infty}$ in the interval of changing *n* from 2,000 to 3,000. This function also behaves in other intervals of changing the number of neutron generations *n*. This dependence is approximated by the function $1.43/n$. For other values of *c* other than $c=c_0$, this dependence does not hold.

In some works (for example, in [3]) the value of the critical region is described as $c \approx c_c + B/L + ...$, where *B* is a constant; *L* is the size of the system. It was noted in [21] that the size of the Bethe lattice is proportional to *lnN*. Since *N=n*, this corresponds to the expression $1.43/n$. It is possible to estimate the time during which the values of $c_0=c_{c\infty}$ reach a given level. For example, the value $10^{-6}$ is achieved in $n=1,43 \cdot 10^6$ generations of neutrons. For thermal neutron reactors, where the average lifetime of a generation taking into account delayed neutrons is $10^{-1}$ s, this time is $1.43 \cdot 10^5$ sec = 1.655 days. For fast neutron reactors, where the average lifetime of a neutron generation is $10^{-4}...10^{-8}$ s, this time decreases by 3...7 orders of magnitude. The considered methods for determining the critical point can be useful during the startup of a reactor or during its transient processes, when the probability value *c* changes due to manipulations with absorber rods.

The rate of change of the function *P(n,c)* depending on *n* can be described by its derivative with respect to *n*, more precisely, by its discrete analogue, the value $P(n+1,c_0)-P(n,c_0)$. Calculations show that this value, taken with a negative sign, for $c_0=c_{c\infty}$ is well described by the dependence $1.43/n^2$. The second derivative, the function $P(n+1,c_0)-2P(n,c_0)+P(n-1,c_0)$, is described by the dependence $2.83/n^3$. The dependence of the k-th derivative with respect to *n* for $c_0=c_{c\infty}$ is proportional to $n^{-(k+1)}$. The behavior of the discrete analogue of the dependence on *c* on the fourth derivative of *P(n,c)* with respect to *n* is shown in Fig. 10. The behavior of the analogue of the second derivative on a smaller scale of change in the value of *c* is shown in Fig. 11. The dependence of the difference between the second derivatives at points c=0.42 and c=0.4 on *n* is shown in Fig. 12.

Similar dependencies are written for derivatives according to (5), (6). Thus, the behavior of the discrete analogue of the first derivative with respect to *n* from the second derivative of *P(n,c)* with respect to *c* (6) is shown in Fig. 14, and the behavior of the dependence of this quantity on *c* for fixed *n*=50 and *n*=15 is shown in Fig. 15, a) and fig. 15, b), respectively. The dependence of *c* on the third derivative with respect to *n* on the second derivative of *P(n, c)* with respect to *c* (6) for *n* = 25 is shown in Fig. 16.

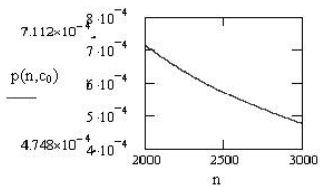

Fig. 9. Dependence of $P(n,c_0)$ on $n$ ($2000<n<3000$) at a fixed value $c=c_0=\overline{V}^{-1}=c_{c\infty}$.

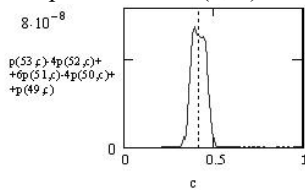

Fig. 10. Behavior of the discrete analogue of the dependence on $c$ on the fourth derivative with respect to $n$ on $P(n,c)$.

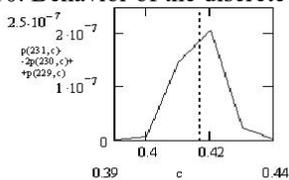

Fig. 11. Behavior of the dependence on $c$ of the analogue of the second derivative with respect to $n$ of $P(n,c)$ ($0.39<c<0.44$).

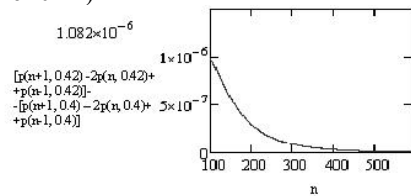

Fig. 12. Dependence on $n$ of the difference between the second derivatives with respect to $n$ of $P(n,c)$ at points $c=0.42$ and $c=0.4$.

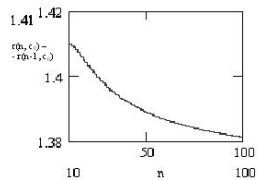

Fig. 13. Behavior of the dependence on $n$ of the discrete analogue of the first derivative with respect to $n$ on the second derivative of $P(n,c)$ with respect to $c$.

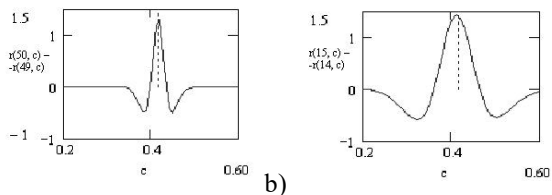

a)     b)

Fig. 14. Behavior of the dependence on $c$ of the discrete analogue of the first derivative with respect to $n$ on the second derivative of $P(n,c)$ with respect to $c$ for fixed $n=50$ (a) and $n=15$ (b).

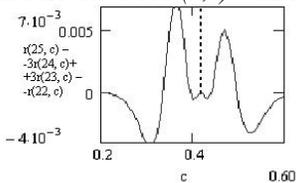

Fig. 15. Dependence of the third derivative with respect to *n* on the second derivative of *P(n,c)* with respect to *c* (6) for *n*= 25.

The dependence on *n* of the third derivative with respect to *n* of *r(n,c)*, the second derivative with respect to *c* of *P(n,c)* is shown in Fig. 15. Position of maxima and minima in Fig. 15 may be related to the boundaries of the critical region. From Fig. 14 shows that the negative peaks for the first derivative with respect to *n* become closer together as *n* increases.

### 3. Conclusion and discussion

The percolation and fractal properties of neutron processes in a nuclear reactor reflect the complex nature of the processes occurring during nuclear fission and the movement of neutrons. The use of exact relations of the theory of percolation obtained for Bethe lattices makes it possible to estimate the temporal behavior of such an important quantity for the practical operation of reactor installations as the multiplication factor. Estimates show that the values of the unit multiplication factor can only be achieved in the practically unrealizable case of an infinite number of neutron generations, corresponding to infinitely large times and infinitely large systems. But for real operating times, it is possible to estimate the times required to achieve very small intervals from the unit value of the multiplication factor.

The results have important implications for the safety of reactor systems with relatively few neutrons, such as during reactor start-up phases or critical assemblies. They should also be useful for a more detailed description of transient processes in the reactor.